\documentclass[runningheads]{llncs}

\usepackage{graphicx}
\usepackage{array}
\usepackage{booktabs}
\usepackage{multicol}
\usepackage{multirow}
\usepackage{makecell}

\usepackage{amsmath}
\usepackage[margin=1.5in]{geometry}

\newcolumntype{N}{>{\centering\arraybackslash}m{.4in}}

\begin{document}
\title{Deep Neural Networks for Query Expansion using Word Embeddings}
%
%
\author{Ayyoob Imani\inst{1} \and
 Amir Vakili\inst{1} \and
Ali Montazer\inst{2} \and Azadeh Shakery\inst{1}}
%
%
 \institute{Tehran University, Tehran, Iran \email{ \{ayyoub.imani, shakery, a\_vakili\} @ut.ac.ir}
 \and University of Massachusetts Amherst, Amherst, USA \email{montazer@umass.edu}}
%
%
\maketitle              
\begin{abstract}
Query expansion is a method for alleviating the vocabulary mismatch problem present in information retrieval tasks. Previous works have shown that terms selected for query expansion by traditional methods such as pseudo-relevance feedback are not always helpful to the retrieval process. In this paper, we show that this is also true for more recently proposed embedding-based query expansion methods. We then introduce an artificial neural network classifier to predict the usefulness of query expansion terms. This classifier uses term word embeddings as inputs. We perform experiments on four TREC newswire and web collections show that using terms selected by the classifier for expansion significantly improves retrieval performance when compared to competitive baselines. The results are also shown to be more robust than the baselines.
\end{abstract}
\section{Introduction}
Query expansion is a method for alleviating the vocabulary mismatch problem present in information retrieval tasks. This is a fundamental problem where users and authors often use different terms describing the same concepts, which in turn harms retrieval performance. In this paper, we aim to distinguish terms helpful to the query expansion process through the use of an artificial neural network classifier.

Various methods for selecting expansion words exist, however, pseudo-relevance feedback (PRF) is the most well-known. PRF assumes that the top retrieved documents for a query will contain terms relevant to the query which can help distinguish other relevant documents from non-relevant documents. However, \cite{cao} showed that not all terms extracted using PRF methods help the retrieval process and many of these terms even have a negative effect. \cite{cao} divided terms into three categories: good, bad and neutral. When used as expansion terms for a query, good terms increase and bad terms decrease retrieval performance. Neutral terms have no effect if added. An ideal query expansion method would add only good terms to the query. 
Therefore, \cite{cao} proposed a supervised learning method for the classification of such terms.

\cite{cao} uses a feature vector to train a classifier for separating good expansion terms from bad ones. This feature vector includes the following features: terms distribution in the feedback documents and terms distribution in the whole collection, co-occurrences of the expansion 
term with the original query 
terms, and proximity of the expansion terms to the query terms.

Another approach for improving the selection of query expansion terms is considering the semantic similarity of the candidate term with the query terms \cite{montazeralghaem2016axiomatic,kuzi,zamani2016embedding,zamani2016estimating,almasri2016comparison}.
As an example, \cite{montazeralghaem2016axiomatic} proposed a semantic similarity constraint for PRF methods and showed that adhering this constraint improves retrieval performance. 

Another method that takes semantic similarity into consideration is using word embeddings to expand queries. \cite{kuzi} aimed to expand queries with terms semantically related to query terms. To this end, they trained word embeddings on document corpora using the Word2Vec Continuous bag of words approach \cite{mikolov}. This technique learns low dimensional vectors for words in a vocabulary based on their co-occurrence with other words within a specified window size. These vectors are both semantic and syntactical representations of their corresponding words. The learning is done unsupervised therefore these vectors are query-independent. \cite{kuzi} then uses these word embeddings to expand queries either by adding terms closest to the centroid of the query word embedding vectors (referred to in following sections as Average Word Embedding or AWE) or combining terms closest to individual query terms in the word embedding space. \cite{zamani2016embedding} has a similar approach for query expansion. It first defines a more discriminative similarity metric than cosine-similarity. Next, it introduces two embedding based query expansion models with different simplification assumptions. One model assumes that query terms are independent of each other (referred to in following sections as Embedding Query Expansion 1 or EQE1), and the other one assumes that the semantic similarity between two given terms is independent of the query. 

In related work \cite{diaz2016query} showed that training local training of word embeddings using retrieved documents improves query expansion effectiveness.  \cite{zheng2015learning} proposed using supervised training and word embeddings to learn term weights to be used in retrieval models such as BM25. 

In this paper, we aim to use pre-trained word embeddings in selecting suitable expansion terms. Furthermore, we propose an artificial neural network (ANN) model for this classification task. We use a siamese neural network architecture inspired by \cite{koch2015siamese} in order to lessen the impact of limited training data. Siamese network architectures have been gaining popularity in recent years in the information retrieval community \cite{yang2017question,dssm,he2016pairwise,severyn2015learning,wang2016compare} and have achieved impressive performance in various tasks. Using the distributed representation of terms, this network learns whether a term is semantically suitable for expanding a query. Our neural network approach intends to go beyond simple vector similarity and learns the latent features present within word embeddings responsible for term effectiveness or ineffectiveness when used for query expansion. In short, the proposed approaches main advantages are that it no longer requires manual feature design and no longer relies on simplistic similarity functions as it uses a trainable classifier.  

We evaluate the effectiveness of our approach on four TREC collections. 
We compare results with traditional approaches and more recent methods. Results show that integrating term classifications using our novel ANN model significantly improves retrieval performance. We also show that our proposed method is more robust compared to the baselines.


The rest of this paper is organized as follows: In section \ref{sec:termGoodnessSec} we discuss \cite{cao}'s method for labeling expansion terms in greater detail, in \ref{sec:modelsec} we introduce our classifier model and explain its integration with the retrieval process, in section \ref{sec:experiment} we present experiments performed and in section \ref{sec:results} we discuss the results obtained from the experiments.

\begin{table}
\caption{Query expansion term statistics for used collections} \label{tab:stats}
\centering
\begin{tabular}{l N N N N N N N N N N}
        \toprule
         \multirow{2}{*}{} & 
         \multicolumn{4}{c}{Embedding-based} &
         \multicolumn{4}{c}{Pseudo Relevance Docs} &
         \multirow{2}{*}{\makecell{QLM \\ (MAP)}}  &
         \multirow{2}{*}{\makecell{QLM \\ +MM \\ (MAP)}} \\ 
         \cmidrule(lr){2-5}\cmidrule(ll){6-9}
                    &  Good (\%)  & Neutral (\%)  & Bad (\%)   & Oracle (MAP)   & Good (\%)   & Neutral (\%)  & Bad (\%)   & Oracle (MAP)    &  & \\
         \midrule
         AP         & 3.8 & 55.5 & 40.5 & 0.2981 & 16.2 & 53.6 & 30.1 & 0.3983 & 0.2206 & 0.2749   \\
         Robust     & 5.6 & 62.4 & 31.9 & 0.3122 & 21.9 & 55.4 & 22.6 & 0.4021 & 0.2176 & 0.2658   \\
         WT2g       & 6.2 & 37.8 & 55.8 & 0.3442 & 15.7 & 61.3 & 22.9 & 0.4383 & 0.2404 & 0.2593  \\
         WT10g      & 3.4 & 75.6 & 20.9 & 0.2410 & 14.1 & 63.6 & 22.2 & 0.2871 & 0.1837 & 0.1902  \\
         \bottomrule
    \end{tabular}
\end{table}

\section{Good, bad, and neutral expansion terms} \label{sec:termGoodnessSec}
In order to identify terms helpful to query expansion, we follow \cite{cao} and divide candidate terms into three classes: good, bad and neutral. For a particular query, a good term will increase retrieval performance and a bad term will decrease it. Neutral terms have no effect. In order to train our classifier, we require a dataset containing query and candidate expansion terms which have been labeled as one of the three classes mentioned.

For creating this dataset, we take each query and first average the embedding vectors of its query terms; this approach has been proposed in \cite{mikolov} and used in other works such as \cite{kuzi,zamani2016embedding}. Then, we use cosine-similarity to find top 1000 terms that are closest to the averaged vector. Finally, to identify the class of a term, we use the method proposed at \cite{cao}). Briefly, we add the expansion to the query and perform retrieval using Query likelihood method; if the mean average precision increases, the term is a good expansion term, but if it decreases, the term is a bad expansion term. If the change is not tangible, the term is neutral.

To examine the impact of selecting good expansion terms, we explore the performance of an oracle retrieval model that expands queries with only good expansion terms. Table \ref{tab:stats} shows the ratio of the classes of terms along with the performance of the oracle retrieval model. The first set of columns depicts statistics for the top 1000 closest terms to the average of query word embeddings, and the second set is for the terms that exist in first 10 pseudo-relevant documents retrieved using the query likelihood method. The last two columns show the performance of the original query likelihood method and the expanded query by the mixture model. The performance is measured using mean average precision for the top 1000 results. We can see that only a small percentage of expansion terms are good expansion terms in all collections, which may explain the slight improvements achieved by query expansion methods using word embeddings compared to the improvements by pseudo-relevance methods. 

\begin{figure}
\centering
\includegraphics[width=\textwidth]{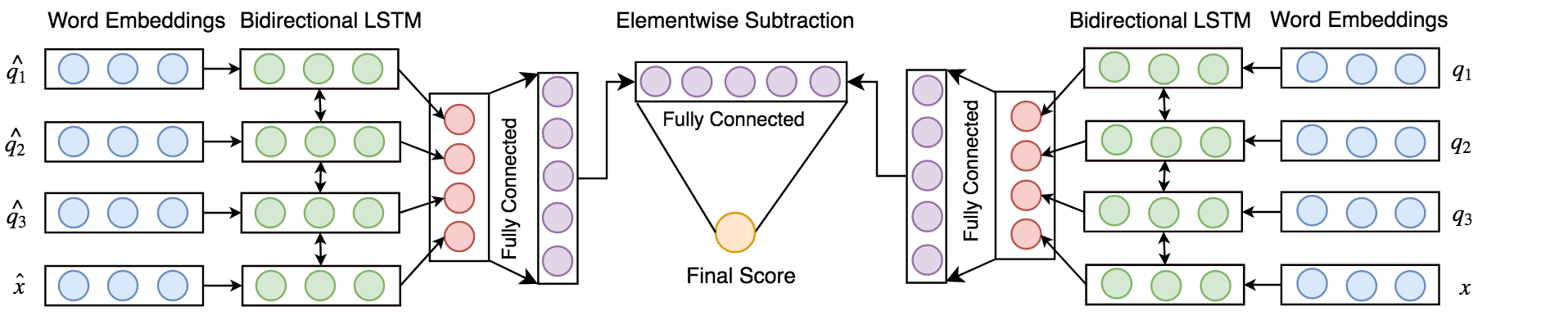}
\caption{Architecture of the proposed siamese network. The architecture consists of two identical models projecting two separate inputs (query and expansion term pair) into a common embedding space and then comparing the two projections to get a final similarity score. This score tells us whether the candidate expansion terms of the two classes belong to the same class (good, bad or neutral).} \label{fig:arch}
\end{figure}

\section{Expansion Term Classification} \label{sec:modelsec}
In this section, we first present the model used for the classification task, then we explain how the classifier results are integrated into the retrieval process.
\subsection{Problem Formulation}
Suppose we have a dataset \(\mathcal{D} = \{(q_i, x_i, l_i)\}_{i=1}^N\) where \(q_i = \{q_{i,1}, \cdots, q_{i,k}\}\) represents a query and \(x_i\) is a candidate expansion term. \(l_i \in \{\text{Good}, \text{Neutral}, \text{Bad}\}\) is the label denoting the expansion terms effectiveness for the query. Our goal is to learn a model \(g(\cdot,\cdot)\) with \(\mathcal{D}\). For any query and candidate expression term pair \((q, x)\), \(g(q, x)\) classifies the expansion term as either good, neutral or bad. 
\subsection{Model Overview}
We propose the deep expansion classifier (DEC) to model \(g(\cdot, \cdot)\). The architecture is depicted in Figure \ref{fig:arch}. A major roadblock when using ANN approaches in information retrieval tasks is the lack of training data. To overcome this, various methods such as using weak supervision for training ranking networks \cite{dehghani2017neural} have been proposed. In this paper, we use the learning technique proposed in \cite{koch2015siamese}. The \textit{siamese network} proposed was designed to overcome the lack of training data by learning whether two samples are of the same class or not rather than directly predicting which class a sample belongs to. Using this technique we can increase available training samples from \(n\) to \(n(n-1)\). In order to determine which class a query and expansion term pair \((\hat{q}, \hat{x})\) belongs to we can use the trained model to compare it to a set of previously classified pairs \((q,x)\), half of which are classified as good. We calculate the probability of a candidate expansion term being good based on the number of times it is put in the same class as a good pair or a different class than a bad pair.


\subsection{Modeling the Query and Candidate Expansion Term Relation}
Given a query \(q\) and a candidate expansion term \(x\), the model connects them as a sequence, then maps the words to their embeddings through a shared look-up table. A bidirectional long short term memory (LSTM) Network is then used to construct new contextual embeddings for each of the terms in the sequence. 
LSTMs \cite{hochreiter1997lstm} are a variant of recurrent neural networks (RNN) designed to overcome the vanishing/exploding gradient problem present in regular RNNs with the usage of memory cells. These cells store information over long histories. The flow of information into and out of these memory cells is controlled by three gates (input, output and forget). Given a sequence of word embeddings including \(k\) query terms and one expansion term \((e_{q,1}, \cdots, e_{q,k},  e_{x})\), an LSTM will output a new sequence where each element captures the contextual information seen before it.


In bidirectional LSTMs the sequence is propagated both forwards and backwards and the two forwards and backwards contextual embeddings are concatenated. The outputs of the bidirectional LSTM layer is fed to a fully connected layer with a softmax activation function. This final layer represents the relation between query terms and the candidate expansion term. 

\subsection{Expansion term classification and Term re-weighting}

The representation built must then be compared to the representation of another query-expansion term pair in order to determine whether the two pairs belong to the same class or not. An element-wise subtraction is performed on the two representations and the result is then fed into a fully connected layer with a softmax activation function which outputs the final score depicting whether the expansion terms belong to the same class.

In order to determine whether an expansion term candidate is good for a query, we feed the pair along with a set of query-expansion term pairs whose classes have been previously determined. We calculate the probability of a candidate expansion term being good as \(P(l=\text{Good}|(q,x)) = \frac{N_g + N_{nb}}{N}\) where \(N\) is the total number of pairs we compare to and \(N_g + N_{nb}\) is the number of times the pair being tested was determined to be in the same class a good pair or in a different class than a bad pair.

Finally, we reweight the expansion term weights obtained using the AWE method by multiplying them by \(P(l=\text{Good}|(q,x))\). The AWE method weights the candidate expansion terms based on the cosine similarity of the expansion term embedding and the query term embeddings centroid. So the final weight for an expansion term will be:
\(
     (1 + \alpha \cdot p(l=\text{Good})) \cdot \delta(\Bar{q},x)
\)
where \(\delta\) is the cosine similarity function and \(\alpha\) is a hyper-parameter.

\section{Experiment} \label{sec:experiment}
For evaluating the proposed methods we use four standard TREC collections: AP (Associated Press 1988-1989), Robust (TREC Robust Track 2004), WT2G and WT10G (TREC Web track 1999, 2000-2001). The first two collections contain news articles and the second two are general web crawls. We used the title of topics as queries. The words are stopped using the inquiry stop word list and no stemming was used.
We used pre-trained word embeddings with a dimension of 200 extracted using the GloVe \cite{pennington2014glove} method on a 6 billion token collection (Wikipedia 2014 dump plus Gigawords 5). 

The parameters were updated by stochastic gradient descent with the Adam algorithm. Hidden size for the bidirectional LSTM is set to 200. The query and expansion term representation vector is set to 400. Our initial learning rate is 0.001 with a mini-batch size of 32. We train models on a single machine. We use k-fold cross validation (k depends on the number of queries available for each dataset) and average the evaluation metrics. 
\subsection{Comparison Approaches}
For evaluation, we only compare to methods that select expansion term candidates based on word embeddings and not other information sources such as the top retrieved documents for each query (PRF). As we use general purpose word embeddings, we also do not compare to methods that train embeddings specifically for query expansion such as \cite{sordoni2014learning,zamani2017relevance}.

For evaluating our proposed method we consider three baselines: (1) the standard query likelihood model using maximum likelihood estimation, (2) AWE where expansion terms closest to the centroid of query term embeddings are selected \cite{kuzi}, and (3) EQE1 where expansion terms are scored by their multiplicative similarity to the query terms  \cite{zamani2016embedding}.

\subsection{Evaluation Metrics}

We use mean average precision (MAP) as our main evaluation metric. We also compare precision for the top 10 retrieved documents (P@10). Statistical significance tests are performed using two-tailed paired t-test at a 95\% confidence level. In order to evaluate the robustness of the proposed methods we use the robustness index (RI) \cite{collins2009reducing} which is defined as \(\frac{N_+-N_-}{|Q|}\) where \(|Q|\) is the number of queries and \(N_+\)/\(N_-\) are the number of queries which have improved/decreased compared to the baseline. 

\begin{table} \label{tab:results}
\caption{Evaluation results on the four datasets. The superscripts 1/2/3 denote that the MAP improvements over QLM/AWE/EQE1 are statistically significant. The highest value in each column is marked in bold. }\label{tab:results}
\centering
\begin{tabular}{lllllllllllll}
        \toprule
         \multirow{2}{*}{} & 
         \multicolumn{3}{c}{AP} & \multicolumn{3}{c}{Robust} & \multicolumn{3}{c}{WT2g} & \multicolumn{3}{c}{WT10g}\\ 
         \cmidrule(lr){2-4}\cmidrule(lr){5-7}\cmidrule(lr){8-10}\cmidrule(lr){11-13}
                    &  MAP  & P@10  & RI    & MAP   & P@10  & RI    & MAP   & P@10  & RI    & MAP   & P@10  & RI \\
         \midrule
         QLM        & 0.2206  & 0.3432 & - & 0.2176 & 0.3856 & - & 0.2404 & 0.4167 & -  & 0.1837 & 0.2420 & - \\
         AWE        & $0.2312^{1}$  & 0.3392 & 0.12 & $0.2230^{1}$ & 0.3899 & 0.10 & $0.2456^{1}$ & 0.4169 & 0.08 & 0.1849 & 0.2399 & 0.10 \\
         EQE1       & $0.2344^{12}$  & $\boldsymbol{0.3442}$ & 0.29 & $0.2278^{12}$ & 0.4016 & 0.25 & $0.2463^{1}$ & 0.4188 & $\boldsymbol{0.17}$ & 0.1867 & 0.2432 & 0.18 \\
         DEC        & $\boldsymbol{0.2403}^{123}$  & 0.3434 & $\boldsymbol{0.31}$ & $\boldsymbol{0.2358}^{123}$ & $\boldsymbol{0.4057}$ & $\boldsymbol{0.31}$ & $\boldsymbol{0.2489}^{1}$ & $\boldsymbol{0.4213}$ & 0.16 & $\boldsymbol{0.1891}^{12}$ & $\boldsymbol{0.2434}$ & $\boldsymbol{0.20}$ \\
         \bottomrule
    \end{tabular}
\end{table}

\section{Results and Discussion} \label{sec:results}
Table \ref{tab:results} presents the performance of the baselines and our proposed method. These methods expand queries with semantically related terms. The results show that the query expansion classifier DEC outperforms all baselines in terms of MAP. For all three embeddings-based methods, the performance gains are more pronounced in the two newswire collections. This may be due to the fact that these collections are more homogeneous than the web corpora. Web corpora will generally be noisier which in turn will affect classifier performance. Another reason could be due to the fact that our word-embeddings (GloVe) were pre-trained on Wikipedia and newswire articles. This would make them more suitable for use in newswire collections. Using word embeddings pre-trained using common crawl data may yield better performance in web corpora.

To summarize, the proposed method outperforms other state-of-the-art methods utilizing word embeddings for query expansion. The results are also more robust than previous approaches. This indicates that for selecting candidate expansion terms, simple similarity functions such as cosine similarity can be improved upon using ANN classifiers. 

\section{Conclusion and Future Work} \label{sec:conclusion}
In this paper, we proposed a neural network architecture for classifying terms based on their effectiveness in query expansion. The neural network uses only pre-trained word embeddings and no manual feature selection or initial retrieval using the query is necessary. We evaluated the proposed methods using four TREC collections. The results showed that the proposed method significantly outperforms other word embedding based approaches and traditional pseudo-relevance feedback. The method is also shown to be more robust than the baselines. 

For future work, one possible method is integrating topic vectors trained using methods such as Latent Dirichlet Allocation into the classification process. Another is using word embeddings trained using methods other than Word2Vec and GloVe (such as Lda2Vec \cite{moody2016mixing} or Paragraph2Vec \cite{le2014distributed}) or training on domain-specific corpora.

%
%
%
\bibliographystyle{splncs04}
\bibliography{samplepaper}
\end{document}